\documentclass[aps, amsfonts, nofootinbib,notitlepage, preprintnumbers]{revtex4-1}
\usepackage{hyperref}
\hypersetup{
colorlinks=true,
linkcolor=red,
citecolor=blue,
urlcolor=blue}
\usepackage{feynmp}
\usepackage{epsf,epsfig}
\usepackage{amscd}
\usepackage{amsmath}
\usepackage{subfig}
\usepackage{graphicx}
\usepackage[countmax]{subfloat}
\input xy
\xyoption{all} 

\newcommand{\figref}[1]{Fig.~\ref{#1}}

\newcommand{\beq}{\begin{equation}}
\newcommand{\eeq}{\end{equation}}
\newcommand{\bea}{\begin{eqnarray}}
\newcommand{\eea}{\end{eqnarray}}

\newcommand{\nn}{\nonumber}

\newcommand{\threeptwo}{${}^3P_2\ $}

\begin{document}

\title{Neutrino emissivity from Goldstone boson decay in magnetized neutron matter} 
\author{Paulo Bedaque\footnote{{\tt bedaque@umd.edu}}} 
\author{Srimoyee Sen\footnote{{\tt srimoyee@umd.edu}}}
\affiliation{Maryland Center for Fundamental Physics,\\ 
Department of Physics,\\ 
University of Maryland, College Park, MD USA}


\begin{abstract}
	Neutron matter at densities somewhat above nuclear densities is believed to be superfluid due to the condensation of neutron pairs in the \threeptwo channel. This condensate breaks rotational symmetry spontaneously and leads to the existence of Goldstone bosons (angulons). We show that  the coupling to magnetic fields mediated by the  magnetic moment of the neutron makes angulons  massive and  capable of decaying into a neutrino-antineutrino pair. We compute the rate for this process and argue they become competitive with other  cooling processes for temperatures around $10^{7} K$ as long as the {\it interior} magnetic field of the star is in the $B\approx 10^{15} G$ range or above.
	\end{abstract}
\maketitle

\section{Introduction}
  The cooling of pulsars, either after their formation or after an accretion episode, is an important probe of their interior. Contrary  to the star's mass and radius relation, determined by  the equation of state of dense matter, cooling process are sensitive to the effective degrees of freedom in the star core.  The temperatures  at which neutron stars are typically found are much smaller than the Fermi energies and are only capable of exciting  the low lying modes. It is the decay and interactions of these modes that control the neutrino emission process leading to loss of energy and cooling of the star. 
  The properties of the low lying degrees of freedom are very sensitive to the  thermodynamic phase realized in the star: the low lying modes of quark matter are different from the ones in neutron matter and both are sensitive to neutron, proton and quark pairing. Therefore, the comparison of neutron star cooling curves  with theoretical predictions based on different models of dense matter is  a way of finding out which thermodynamic phase is actually realized at the core os the star and can even distinguish between two phases whose equations of state are very similar.   The recent observations of neutron stars with mass above two solar masses \cite{Antoniadis:2013pzd},  \cite{Demorest:2010bx} strongly suggest that "exotic" phases -- those made up of other particles besides neutron, protons and electrons --   are not present in neutron stars. It is then natural to focus the attention onto non-exotic phases. The theoretical expectation, based on vacuum phase shifts and model calculations, is that neutron matter is paired at the relevant densities, with the pairing occurring in the ${}^1S_0$ channel at lower densities and on the \threeptwo channel at higher densities. The protons, due to their lower density, are expected to be paired in the ${}^1S_0$ channel. The condensation of neutrons in the \threeptwo channel leads to the appearance of a unique class of low energy excitations. As the condensate is a spin 2 object, its orientation in space defines a special frame and breaks  rotational symmetry spontaneously. As a consequence, one expects Goldstone bosons (named ``angulons"), which are ungapped scalar excitations, to exist in the  \threeptwo  phase. This observation was made some time ago in \cite{Bedaque:2003wj} where one of the variants of the \threeptwo  phase was considered. 
The recent observation of rapid cooling of the neutron star Cassiopeia A   \cite{Heinke:2010cr} has renewed interest in the phenomenon of neutron pairing and condensation as the phenomenon was explained \cite{Sedrakian:2013pva} \cite{Shternin:2010qi} \cite{Blaschke:2011gc} as a result of neutrino emission in the process of  \threeptwo  Cooper pair breaking and formation  (PBF) \cite{Flowers:1976ux}  \cite{Leinson:2011jr}. The PBF process is only effective at temperatures close to the critical temperature where unpaired neutrons exist in substantial numbers. At lower temperatures, processes involving angulons are likely to dominate. In  \cite{Bedaque:2003wj}  an estimate was made for the bremsstrahlung of neutrino pairs following a angulon-angulon collisions. The emissivity was found to be proportional to $T^9$ and small for most relevant densities and temperatures. Angulon decay into neutrino pairs is kinematically forbidden as the angulon momentum is space-like. The main point we make of the present paper is that the dispersion relation of the angulons is changed  by the presence of strong magnetic fields and one of them develops a gap of the order of $eB/M$ ($B$ is the magnetic field, $e$ the electron charge and $M$ the neutron mass, corrected by Fermi liquid effects). The gapped angulon is then kinematically allowed to decay into a neutrino pair. 

In order to proceed with the calculation of the emissivity due to magnetic field catalyzed  angulon decay we first review some angulon properties derived in \cite{Bedaque:2012bs} where the low-energy effective theory for the \threeptwo phase was developed.
The core of a neutron star is expected to have densities above the nuclear matter saturation density at a temperature well below the Fermi momentum of the neutrons and the attractive force between neutrons near the Fermi surface leads to Cooper pair formation. At moderate densities, $3\times10^{11}\text{g/cm}^3\leq\rho\leq 10^{14}\text{g/cm}^3$ the neutrons form s-wave Cooper pairs  while further inside the core, at even higher densities ranging from $1.5\times10^{14}\text{g/cm}^3$ to $10^{15}\text{g/cm}^3$,  neutrons undergo triplet $(^{3}P_{2})$ pairing due to short range spin orbit interaction \cite{Muzikar:1980as}. The order parameter for the  triplet condensed phase is given by
\bea\label{eq:condensate}
\langle n^T \sigma_2 \sigma_i \overleftrightarrow{\nabla} _j n\rangle =\Delta_{ij}e^{i\alpha},
\eea where $\Delta_{ij}$ is a symmetric traceless matrix \cite{Sauls:1982ie, PhysRevD.17.1524}. Here, $n$ is the neutron field, $\sigma$ are Pauli spin matrices and $\alpha$ is an arbitrary phase. Different forms of $\Delta_{ij}$ break rotational symmetry in different ways and lead to different angulon properties. Near the critical temperature, where Guinzburg-Landau arguments are valid, the condensate in the energetically favorable ground state is \cite{PhysRevD.17.1524}
\bea
\label{eq:r}
\bar\Delta=\Delta_0 \begin{pmatrix}
1 & 0 & 0\\
0 & r & 0\\
0 & 0 & -1-r
\end{pmatrix},
\eea  with $r=-1/2$ ( assuming certain parameters  are not too different from the BCS values) and it has been argued this pattern is stable as the temperature is lowered \cite{Khodel:1998hn}. It will be an assumption of our calculation that the condensate has the form in eq.~\ref{eq:r}; we will comment in the conclusion how our results would change if a different form of \threeptwo   pairing were to occur. The presence of  a magnetic field has two effects on the condensate. First, it becomes energetically favorable for the direction  with the eigenvalue $1$ in eq.~\ref{eq:r}, which otherwise would be arbitrary,  to align with the magnetic field \cite{Muzikar:1980as}. Second, the value of $r$ changes slightly to $r=-1/2+C B^2 \approx -1/2+(0.017 B_{15})^2$, where $C$ is a combination of parameters of the Guinzburg-Landau free energy and $B_{15}=B/(10^{15} G)$ \cite{Muzikar:1980as}. We will neglect the change in $r$ due to the magnetic field which is a good approximation for $B\lesssim 10^{17} G$. 

The condensate in eq.~\ref{eq:condensate} breaks spontaneously the rotation and phase invariance symmetries $SO(3)\times U(1)$ down to the subgroup $O(2)\times \mathbf{Z}_2$ composed of rotations around the z-axis (direction of the magnetic field) and rotations of the phase of the neutron field by $\pi$ which leave the condensate in eq.~\ref{eq:condensate} invariant. As a consequence we expect two Goldstone bosons corresponding to rotations of the condensate around the $x$ and $y$ axis. Since, by rotational symmetry, long wavelength oscillations of the condensate in these directions cost no energy, the quantized modes will be ungapped, in accordance with the Goldstone theorem.  These oscillations can be parametrized by the fields $\alpha_1$ and $\alpha_2$ defined by
\beq
\Delta(x) =  e^{i(J_1 \alpha_1 + J_2 \alpha_2)/f}\bar\Delta e^{-i(J_1 \alpha_1 + J_2 \alpha_2)/f},
\eeq where $J_1, J_2$ are the $3\times 3$ matrix generating rotations around the $x$ and $y$ axis. The effective theory for the angulons was derived in \cite{Bedaque:2012bs} under mild assumptions. Due to the lack of rotational symmetry the explicit expression of even the lowest order terms ( in powers of $\alpha_i$ and in the number of derivatives), is unenlightening and can be found in \cite{Bedaque:2012bs}.

For our purposes only two terms are important. The first is the leading (two derivatives)  term quadratic in $\alpha_i$ in the presence of a magnetic field which, in momentum space reads

\bea\label{eq:S}
S &=& 
\int \frac{d^4p}{(2\pi)^4}\begin{pmatrix}
\alpha_1(p) &
\alpha_2(p)
\end{pmatrix}
\begin{pmatrix}
a p_0^2+v_F^2(dp_x^2+cp_y^2+bp_z^2) & ev_F^2p_xp_y-i\frac{eg_NBp_0}{2M}\\
ev_F^2p_xp_y-i\frac{eg_NBp_0}{2M} & a p_0^2+v_F^2(dp_x^2+cp_y^2+bp_z^2)
\end{pmatrix}
\begin{pmatrix}
\alpha_1(-p) \\
\alpha_2(-p)
\end{pmatrix}
\eea 
where $a$, $b$, $c$, $d$, $e$ are given by \cite{Bedaque:2012bs}
\bea
a &=& 3+\frac{\pi}{\sqrt{3}} \approx 4.81 ,\qquad b = -\frac{3}{2}+\frac{\pi}{9\sqrt{3}} \approx -1.30,\qquad c = -\frac{4\pi}{3\sqrt{3}}\approx -2.42,\\
d &= &-\frac{3}{2}+\frac{2\pi}{9\sqrt{3}}\approx -1.10,\qquad e = \frac{3}{2}-\frac{14\pi}{9\sqrt{3}}\approx -1.32,\\
\eea $v_F$ is the Fermi velocity of the neutrons, $B$ is the magnetic field which points along the $z$ direction, $e$ is the charge of an electron and $g_N$ and $M$ are the magnetic moment and mass of the neutrons respectively, including Fermi liquid corrections. The symbol $e$ has been used to denote both a low energy constant and the electric charge. However, it will always be clear from the context which of the two quantities the symbol stands for.
The coupling of the angulons to the magnetic field can be obtained from the results in  \cite{Bedaque:2012bs} by noticing that magnetic fields couples to neutron through their magnetic moment:
\beq
\mathcal{L}_{B-\alpha} = \frac{e g}{2M} n^\dagger \mathbf{S}.\mathbf{B} n,
\eeq where $g=-1.913$ is the neutron anomalous magnetic moment ( in unit of the nuclear magneton). This coupling has the same form as the angulon coupling to the spatial part of the $Z_0$ boson worked out in \cite{Bedaque:2012bs}, from which we can read off the terms proportional to $B$ in eq.~\ref{eq:S}.

\begin{figure} 
\includegraphics[width=.4\textwidth]{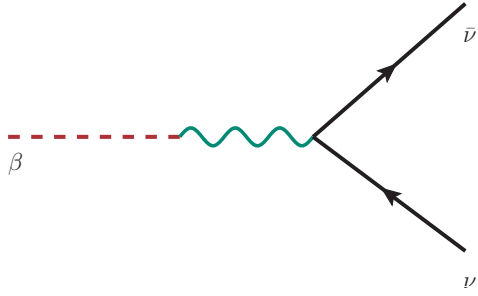}
\caption{
Feynman diagram  showing the massive angulon (dashed line) decay into a neutrino pair (solid line). The wavy line represents a $Z_0$
}\label{fig:f1}
\end{figure}

In order to diagonalize the action we introduce  new fields $\beta_1$ and $\beta_2$ 
\begin{equation}
\sqrt{a}\begin{pmatrix}\alpha_1 \\ \alpha_2 \end{pmatrix} = \begin{pmatrix}c_{11} & c_{12} \\ c_{21} & c_{22}\end{pmatrix} \begin{pmatrix}\beta_1 \\ \beta_2 \end{pmatrix}
\end{equation}
in terms of which the quadratic part of the action reads
\bea
S &=& 
\int \frac{d^4p}{(2\pi)^4}\begin{pmatrix}
\beta_1(p) &
\beta_2(p)
\end{pmatrix}
\begin{pmatrix}
(p_0-\xi_1)(p_0-\xi_2) & 0\\
0 & (p_0+\xi_1)(p_0+\xi_2)
\end{pmatrix}
\begin{pmatrix}
\beta_1(-p) \\
\beta_2(-p)
\end{pmatrix}
\eea 
where, $\xi_1$ and $\xi_2$ determine the dispersion relation of the modes:

\bea
\xi_1 &=& A+\sqrt{A-B}, \\
\xi_2 &=& A-\sqrt{A-B}
\eea where 
\bea
A=\frac{1}{2}\left(  \frac{e g_N B}{2Ma} \right)^2-\frac{c+d}{2a}v_F^2(p_x^2+p_y^2)-\frac{b}{a}v_F^2p_z^2
\eea
and
\bea
B=\frac{cd}{a^2}v_F^4(p_x^4+p_y^4)+\frac{bc+bd}{a^2}v_F^4(p_x^2+p_y^2)p_z^2+\frac{b^2}{a^2}v_F^4p_z^4
+\frac{c^2}{a^2}v_F^4p_x^2p_y^2+\frac{d^2}{a^2}v_F^2p_x^2p_y^2-\frac{e^2}{a^2}v_F^2p_x^2p_y^2.\eea

 We notice that the presence of the magnetic field turned one of the Goldstone bosons into a massive mode while the remaining massless mode has now a quadratic dispersion relation at small momenta.
This is in accord to the generalized Goldstone theorem valid in the absence of Lorentz symmetry  \cite{Nielsen:1975hm}.

The second relevant term of the effective action is the coupling of the angulons to the electroweak $Z_0$ gauge boson  \cite{Bedaque:2012bs}:

\bea
\mathcal{L}=C_A 9f(Z_2^0 \partial_0\alpha_2 - Z_1^0\partial_0\alpha_1)
\eea
where $f^2=\frac{Mk_F}{6\pi^2}$, $k_F$ is the neutron Fermi momentum, $C_A^2=\tilde{C_A}^2\frac{G_F M_Z^2}{2\sqrt{2}}$ with $\tilde{C_A}\sim 1.1\pm0.15$, $G_F$ the Fermi constant and $M_Z$ the $Z_0$ boson mass.

Finally, the coupling between the gauge boson and neutrinos is well know \cite{Peskin}:
\bea
\mathcal{L}_{Z-\nu}=\frac{g Z_{\mu}}{\cos\theta_W}\left(\frac{1}{4}\bar{\nu}\Gamma^{\mu}(1-\Gamma^5)\nu\right).
\eea

\begin{figure}
\includegraphics[width=.4\textwidth]{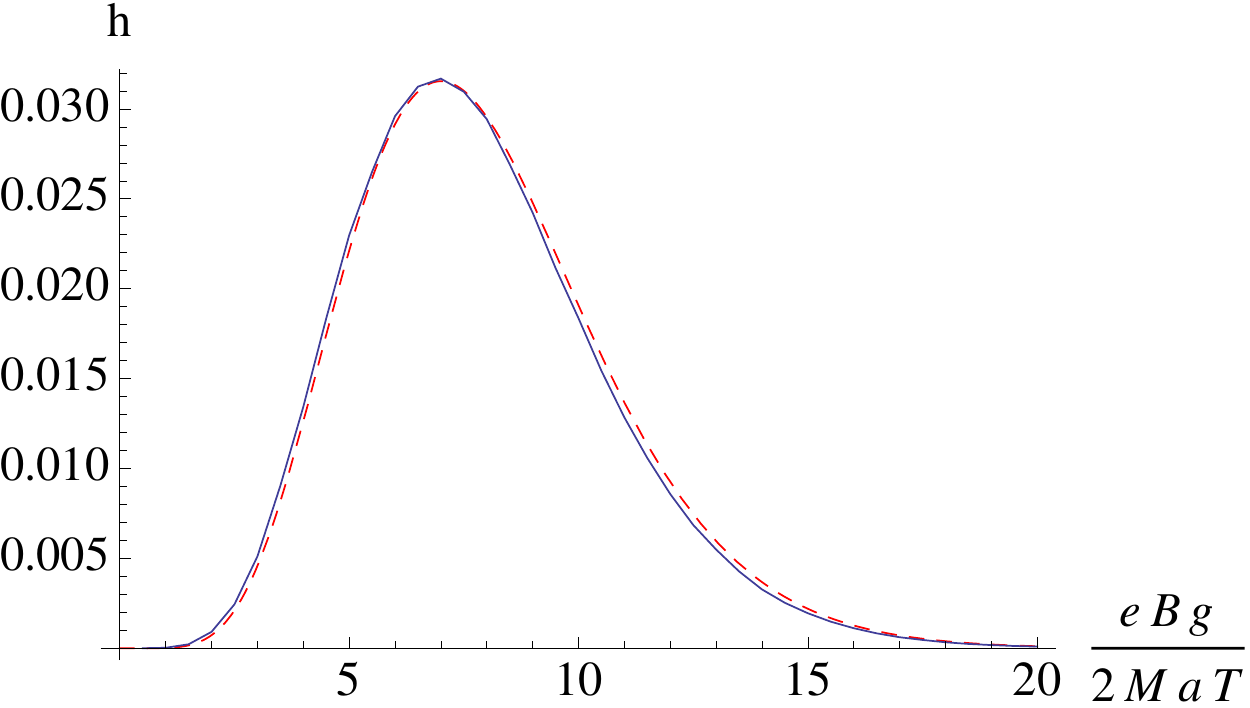}}
\caption{
$h(x)$ as a function  of $x=\frac{eg_N B}{2MaT}$ (solid line)  and its analytic approximation $h(x) \approx 0.000042\ x^7 e^{-x}$ (dashed line).
}{\label{fig:f2}
\end{figure}

\section{Emissivity}\label{sec:three}

The tree level contribution to the massive angulon decay is given by the diagram in 
fig.~ \ref{fig:f1} . 
Using appropriate normalization for the states, the amplitude for this process can be written as
\bea
\label{A}
A =\left(\overline{u}_s(p)\frac{\gamma^1(1-\gamma^5)}{2}v_t(p')c_{11}-\overline{u}_s(p)\frac{\gamma^2(1-\gamma^5)}{2}v_t(p')c_{21}\right)\frac{C_A9fg}{2\sqrt{a}\cos\theta_W}\frac{k_0}{M_Z^2}\frac{(2\pi)^4\delta^4(p+p'-k)}{V^{3/2}\sqrt{2\xi_k2w_p2w_{p'}}}
\eea
where $p$ and $p'$ are the momenta of the outgoing neutrinos and $k$ is the momentum of the angulon $\beta_1$. The on-shell conditions for the external legs are 
\bea
k_0=\xi_1
\nn\\
p_0=w_p=|p|
\nn\\
p'_0=w_p'=|p'|
\eea
as the neutrinos are taken to be massless.

\begin{figure}
\includegraphics[width=.4\textwidth]{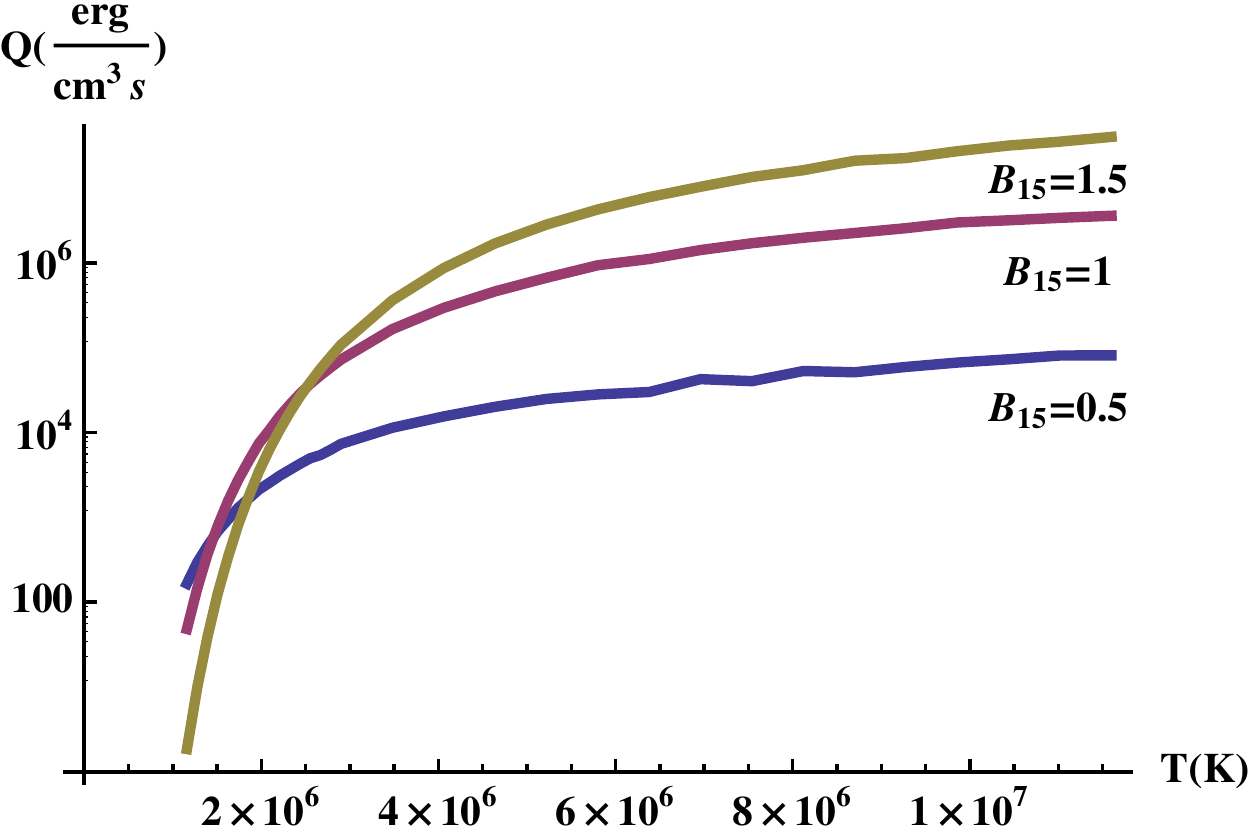}
\caption{ 
Neutrino emissivity as a function of temperature for different magnetic fields. 
}\label{fig:f3}
\end{figure}

 The decay rate is defined by
\bea
\label{gamma1}
\Gamma = N    \mathop{\sum_{\text{neutrino momenta}}}_{\text{and heliticities}}  \frac{|A|^2}{\tau}
\eea
where $\tau$ is time over which the interaction is on (to be taken to infinity at the end of the computation) and $N=3$ is the number of neutrino flavors. The square of the amplitude brings two four dimensional $\delta$-functions and one of them can be replaced by $V\tau$ ($V$ is the volume of the space). After performing the phase space integral over one of the outgoing neutrino momentum we end up with
\bea
\label{gamma2}
\Gamma= N\int\frac{d^3p}{(2\pi)^3}\frac{2\pi\delta(p_0+p_0'-k_0)}{2\xi_k2w_p2w_{p'}}2(P_1|c_{11}|^2+P_2|c_{11}|^2-P_{12}c_{11}c_{21}-P_{12}^*c_{11}^*c_{21})\frac{C_A9fg}{2\sqrt{a}\cos\theta_W}\frac{k_0}{M_Z^2}
\eea
where, 
\bea
\label{P}
P_1=p_0p_0'+p_1p_1'-p_2p_2'-p_3p_3' 
\nn\\
P_2=p_0p_0'+p_2p_2'-p_1p_1'-p_3p_3' 
\nn\\
P_{12}=p_1p_2'+p_2p_1'+i(p_0p_3'-p_3p_0') 
\eea
with $\overrightarrow{p'}=\overrightarrow{k}-\overrightarrow{p}$. The emissivity (the amount of energy emitted per unit of volume and time) is then  given by,
\bea
\label{Q}
Q=\int \frac{d^3k}{(2\pi)^3}\Gamma\frac{k_0}{e^{\beta k_0}-1}
\eea
where $\beta$ is the inverse temperature. 
In order to write $Q$ as a dimensionless integral, we normalize all our symbols with respect to $T^{\lambda}$ where $\lambda$ is the mass dimension of the quantity the symbol stands for. In terms of the new dimensionless symbols, which we distinguish from the unnormalized ones by a tilde, the emissivity can be expressed as
\begin{align}
\begin{split}
Q&= N\int T^5 \frac{d^3\tilde{k}d^{3}\tilde{p}}{(2\pi)^6}(2\pi)\delta\left(\tilde{p_0}+\tilde{p_0'}-\tilde{k_0}\right)\\ 
&\qquad\quad
2(\tilde{P_1}|c_{11}|^2+\tilde{P_2}|c_{11}|^2-\tilde{P_{12}}c_{11}c_{21}-\tilde{P_{12}}^*c_{11}^*c_{21})\left(\frac{\tilde{k_0}}{\tilde{M_z}^2}\right)^2\frac{1}{2\tilde{w_k}2\tilde{w_p}2\tilde{w_p'}}\left(\frac{C_A9\tilde{f}g}{2\sqrt{a}\cos(\theta_W)}\right)^2\frac{\tilde{k_0}}{e^{\tilde{k_0}-1}}.\end{split}\label{QQ}\end{align} The factor of $\left(\frac{C_A9\tilde{f}g}{2\sqrt{a}\cos(\theta_W)}\right)^2\frac{1}{\tilde{M_z}^4}$ can be written as $\tilde{C_A}^2\frac{81}{12a\pi^2}\tilde{G_F}^2\tilde{M}\tilde{k_F}$ and plugging this back in \eqref{QQ} we get,
\begin{align}
\begin{split}
Q&= N \tilde{C_A}^2\frac{81}{12a\pi^2}\tilde{G_F}^2\tilde{M}\tilde{k_F}\int T^5 \frac{d^3\tilde{k}d^{3}\tilde{p}}{(2\pi)^6}(2\pi)\delta\left(\tilde{p_0}+\tilde{p_0'}-\tilde{k_0}\right)\\ 
&\qquad\quad
2(\tilde{P_1}|c_{11}|^2+\tilde{P_2}|c_{11}|^2-\tilde{P_{12}}c_{11}c_{21}-\tilde{P_{12}}^*c_{11}^*c_{21})\frac{1}{2\tilde{w_k}2\tilde{w_p}2\tilde{w_p'}}\frac{\tilde{k_0}^3}{e^{\tilde{k_0}-1}}.\end{split}\label{QQ1}\\
&= G_F^2Mk_FT^7h\left( \frac{e g_N B}{2 a M T} \right)&
\end{align}
where $h(x)$ is given by the dimensionless integral 
\begin{align}
\begin{split}
h&= N \tilde{C_A}^2\frac{81}{12a\pi^2}\int \frac{d^3\tilde{k}d^{3}\tilde{p}}{(2\pi)^6}(2\pi)\delta\left(\tilde{p_0}+\tilde{p_0'}-\tilde{k_0}\right)\\ 
&\qquad\quad
2(\tilde{P_1}|c_{11}|^2+\tilde{P_2}|c_{11}|^2-\tilde{P_{12}}c_{11}c_{21}-\tilde{P_{12}}^*c_{11}^*c_{21})\frac{1}{2\tilde{w_k}2\tilde{w_p}2\tilde{w_p'}}\frac{\tilde{k_0}^3}{e^{\tilde{k_0}-1}},\end{split}\label{QQ2}\end{align}
which is a function of the dimensionless quantity $x=\frac{Beg_N}{2MaT}$ alone. We compute the integral in equation \ref{QQ2} numerically and plot it as a function of $\frac{Beg_N}{2MaT}$ in Fig. \ref{fig:f2}. 
It turns out that the function $h(x)$ is very well approximated by
\beq\label{eq:hanalytic}
h(x) \approx 0.000042\ x^7 e^{-x}
\eeq
Fig. \ref{fig:f3} is a plot of corresponding neutrino emissivity as a function of temperature for three different magnetic fields.

\section{Discussion}

Our main result is the neutrino emissivity due to the decay of angulons in the presence of a magnetic field as shown in eq.~\ref{Qang} and depicted in \figref{fig:f3}. We provided a simple, precise analytic form in eqs.~\ref{Qang} and \ref{eq:hanalytic}.
The calculation involved a number of approximations, all well controlled and precise enough for the application to the cooling of neutron stars. The first is in the derivation of the parameters of the effective theory whose validity is discussed at length in \cite{Bedaque:2012bs}. Higher orders in the low momentum expansion, either from terms in the effective theory with more terms or from loops are suppressed by factors of 
$(T/\Delta_0)^2$ and are small at temperatures well below the critical one. Effects arising from  a possible finite temperature mass the angulons may acquire \cite{Leinson:2012pn} belongs to this category. In addition, we use $r=-1/2$ for our calculations. A different value of $r$ would mean a different pattern of symmetry breaking. Generically, the rotation group would be broken down to the discrete subgroup $(\mathbf{Z}_2)^3$ of inversion along the principal axis of $\Delta_{ij}$ and three Goldstone bosons would exist. However, only one of them would acquire a mass due to the interaction with the magnetic field and our calculation would be changed just by a small shift of the angulon mass. For magnetic fields in excess of $B\approx 10^{17} G $ the phase with $r=-1$ is expected to be favored \cite{Muzikar:1980as}. This phase is qualitatively  distinct from the other because the neutron are gapless along a certain direction in space. Those ungapped neutron may undergo beta decay and provide a source of neutrinos suppressed only by the restricted phase space of ungapped neutrons. Our calculation would still stand for the angulon part but, in this case, it has to be supplemented by the ungapped neutron part.

The emissivity rates from PBF and angulon decay processes are not to be directly compared. This is because the PBF is, at any given time, effective only on a shell of the star where the temperature is near the critical temperature (which is density dependent). On the other hand, due to the proton superconductivity, magnetic fields are believed to be confined to flux tubes and thus angulons can decay  only inside the flux tubes or in their immediate vicinity. 
The comparison between the angulon and PBF emissivities is further complicated by the uncertainty on the value of the gap (which affects the PBF primarily) and on the value of the magnetic fields in the core (which affect the angulon rate). 
Still, it is instructive to look at their relative numerical values.
The emissivity of neutrinos in PBF processes is given by \cite{Flowers:1976ux}\cite{Leinson:2011jr}
\bea
Q_{PBF}=\frac{4G_F^2Mk_F}{15\pi^5}T^7NF\left(\frac{\Delta}{T}\right)
\eea
where $F$ is a function of ratio of the gap $\Delta$ to temperature peaking at $T\sim \Delta$ and decaying exponentially at smaller temperatures. On the other hand the massive angulon decay gives rise to an emissivity equal to
\beq
\label{Qang}
Q_{ang} = G_F^2 M k_F T^7 h\left(\frac{Beg_N}{2aMT}\right),
\eeq where the function $h(x)$ peaks at $x\sim 7$ and decays exponentially at larger values of $x$  and as $\sim x^7$ at small $x$. For temperatures near the gap value $T\sim \Delta_0$ the PBF process is much larger than the angulon emissivity (assuming $\Delta_0 \gg eB/M$). At lower temperatures, around $T\sim eB_{15}/M \approx 3\times 10^7 K B_{15}$, the angulon emissivity is larger than the one from PBF. For temperatures smaller than that, the angulon process still dominates but the phenomenological interest of these rates is small as it is difficult to observe stars so cold. 

Since the angulon decay process can occur only in regions of high magnetic field it is important to have an estimate of the volume fraction of the star that are close enough to magnetic flux tubes.
 Each flux tube carries a flux quantum equal to $\Phi_0=\pi/e$. Assuming a dipole form for the magnetic field inside of the star, the total flux crossing the star is $\Phi=\pi R_{star}^2 B_{star}$ (where $R_{star}$ and $B_{star}$ are the radius and average magnetic field in the interior of the star).
The number of flux tubes then will be of the order of $N \approx \Phi/\Phi_0 \approx R_{star}^2 e B_{star}$ and they will be separated by an average distance of $L\approx \sqrt{\pi R_{star}^2/N}\approx \sqrt{\pi/(e B_{star})}$. The magnetic fields extends around a flux tube to a distance of the order of the penetration length $\lambda = \sqrt{m/(4\pi \alpha n_p)}$, where $n_p$ is the proton density. Thus, the fraction of the star volume with sizable magnetic fields is of the order of $(\lambda/L)^2 \approx 0.04 B_{15} (0.1 n_0)/n_p$, where $B_{15}=B/10^{15} G$, $n_p$ is the proton density and $n_0=0.16 fm^{-3}$ the nuclear saturation density. Clearly only in stars with very large magnetic fields the angulon decay mechanism may be relevant. Magnetars form a class of neutron stars where fields of this order are known to exist but ordinary neutron stars, with much smaller long range magnetic fields may have magnetic fields  of this order in their interior. It is clear, however, that a proper assessment of the angulon decay mechanism on the cooling curves can only be done with a realistic cooling code.

\section{Acknowledgements}
This work was supported by the U.S. Department of Energy through grant number DEFG02-93ER-40762.

\bibliographystyle{unsrt}
\bibliography{angulon}
\end{document}